\documentclass[ prl,10pt,apshttps://doi.org/10.1016/0375-9601(84)90816-8
]{revtex4-2}

\usepackage[utf8]{inputenc}
\usepackage[english]{babel}
\usepackage{indentfirst}
\usepackage[final]{showkeys}
\usepackage{graphicx}
\usepackage{subfig}
\usepackage{amssymb}
\usepackage{latexsym}
\usepackage{amsthm}
\usepackage{amsmath}
\usepackage{amsfonts}
\usepackage{mathtools}
\usepackage{pgf,tikz}
\usepackage{mathrsfs}
\usepackage{epstopdf}
\usetikzlibrary{arrows}
\usepackage{enumerate}
\usepackage{epigraph}
\usepackage[normalem]{ulem}
\usepackage[a4paper, left=1.5cm, right=1.5cm, top=2cm, bottom=2cm]{geometry}
\usepackage{floatpag}
\DeclareUnicodeCharacter{2212}{-}
\DeclareUnicodeCharacter{03B4}{-}

\newcommand{\me}[1]{\left\langle #1 \right\rangle }

\begin{document}

\title{%Deformation of
  %topological %universality
  %phase transitions by
  Berezinskii-Kosterlitz-Thouless phase transitions with long-range couplings}

\author{Guido Giachetti}
\email{ggiachet@sissa.it}
\affiliation{SISSA and INFN Sezione di Trieste, Via Bonomea 265, I-34136 Trieste, Italy}
\author{Nicol\`o Defenu}
%\email{ndefenu@phys.ethz.ch}
\affiliation{Institute for Theoretical Physics, ETH Z$\ddot{u}$rich, Wolfgang-Pauli-Str. 27, 8093 Z$\ddot{u}$rich, Switzerland}
\author{Stefano Ruffo}
\affiliation{SISSA and INFN Sezione di Trieste, Via Bonomea 265, I-34136 Trieste, Italy}
\affiliation{Istituto dei Sistemi Complessi, Consiglio Nazionale delle Ricerche, Via Madonna del Piano 10, I-50019 Sesto Fiorentino, Italy}
\author{Andrea Trombettoni}
\affiliation{Department of Physics, University of Trieste, Strada Costiera 11, I-34151 Trieste, Italy}
\affiliation{SISSA and INFN Sezione di Trieste, Via Bonomea 265, I-34136 Trieste, Italy}
\affiliation{CNR-IOM DEMOCRITOS Simulation Center, Via Bonomea 265, I-34136 Trieste, Italy}

%\author[1]{Guido Giachetti}
%\author[2]{Nicol\`o Defenu}
%\author[1,5]{Stefano Ruffo}
%\author[1,3,4]{Andrea Trombettoni}
%\affil[1]{SISSA and INFN Sezione di Trieste, Via Bonomea 265, I-34136 Trieste, Italy}
%\affil[2]{Institute for Theoretical Physics, ETH Z$\ddot{u}$rich, Wolfgang-Pauli-Str. 27, 8093 Z$\ddot{u}$rich, Switzerland}
%\affil[3]{Department of Physics, University of Trieste, Strada Costiera 11, I-34151 Trieste, Italy}
%\affil[4]{CNR-IOM DEMOCRITOS Simulation Center, Via Bonomea 265, I-34136 Trieste, Italy}
%\affil[5]{Istituto dei Sistemi Complessi, Consiglio Nazionale delle Ricerche, Via Madonna del Piano 10, I-50019 Sesto Fiorentino, Italy}

\begin{abstract}
  The Berezinskii-Kostelitz-Thouless (BKT) transition is the paradigmatic example of a topological phase transition without symmetry-breaking, where a quasi-ordered phase, characterized by a power law scaling of the correlation functions
  at low temperature, is disrupted by the proliferation of topological excitations
  above the critical temperature $T_{\rm BKT}$. 
  In this letter, we consider the effect of long-range 
  decaying couplings $\sim r^{-2-\sigma}$ on this phenomenon. After pointing out the relevance of this non trivial problem, we discuss the  phase diagram, which is far richer than the corresponding short-range one. It features --
  for $7/4<\sigma<2$ -- a quasi ordered phase 
  in a finite temperature range
  $T_c < T < T_{\rm BKT}$, which occurs between a symmetry broken
  phase for $T<T_c$ and a disordered phase for
  $T>T_{\rm BKT}$. The transition temperature $T_c$ displays
  unique universal features quite different from those of the traditional, short-range XY model.
  Given the universal nature
  of our findings, they may be observed in current experimental realizations in $2D$ atomic,
  molecular and optical quantum systems.
\end{abstract}
\maketitle

\section{Introduction}
\noindent 
Two-dimensional interacting systems are well known not to display conventional symmetry breaking transitions at finite temperature, due to the Hohenberg-Mermin-Wagner theorem \cite{zinnjustin}. Yet, a phase transition may appear driven by topological defects, according to the celebrated Berezinskii-Kosterlitz-Thouless (BKT) mechanism \cite{kosterlitz2017nobel}. In the presence of long-range interactions the Hohenberg-Mermin-Wagner theorem no longer holds and local order parameters, such as the magnetization \cite{kunz1976first}, may have a non-zero expectation value. The general question addressed by this Letter is the fate of the BKT transition when the range of the interactions is increased. The Sak's criterion \cite{sak1973recursion} provides an argument for understanding whether the long-range, power law coupling $\sim 1/r^{d+\sigma}$ in the classical $O(N)$ model affects criticality. It can be formulated as follows: at low momenta the short-range (SR) and long-range (LR) critical two-points functions behave as 
\begin{equation}
    p^{-2+\eta_{\rm sr}} \hspace{0.5cm} \text{vs} \hspace{0.5cm} p^{-\sigma}
\end{equation} 
respectively, where $\eta_{\rm sr}$ is the anomalous dimension of the SR limit. Therefore, one can define a critical value of the range of the interactions, $\sigma_* = 2 - \eta_{\rm sr}$, such that, for $\sigma> \sigma^{*}$, the critical behavior is not affected by LR. The validity of Sak's criterion for the classical $O(N)$ models has been the subject of considerable scrutiny. Indeed, numerical evidences supporting (or rejecting) the Sak's result are notoriously hard to obtain\,\cite{luijten2002boundary,blanchard2013influence,grassberger2013two}. Intense theoretical investigations both via MC simulations\,\cite{luijten2002boundary,angelini2014relations,horita2017upper}, renormalization group (RG) theory\,\cite{brezin2014critical,defenu2015fixed,defenu2016anisotropic} and conformal bootstrap\,\cite{behan2017long} appeared all to confirm the validity of Sak's conjecture for the LR-SR crossover so that it is fair to conclude that the criterion has been a useful tool to understand the critical behaviour of LR interacting systems\,\cite{Luijten,Defenu2020review, dutta2001phase, defenu2017criticality}.
The criterion, is believed to apply to all symmetry breaking transition in $d\geq 2$. The status of the $d=2$ classical XY model is rather different, and only few results (later commented) are known. The main reasons are \\
\indent 
{\it i)} The Sak criterion cannot be straightforwardly applied, since in the SR limit the critical behavior is not described by a single RG fixed point, but rather by a whole line of fixed points with a temperature-dependent exponent $\eta_{\rm sr}$. \\ 
\indent
{\it ii)} Numerically, the large number of non-vanishing couplings, coming form the LR nature of the interaction, along with the logarithmic scaling typical of $2D$ systems (the so-called ``Texas state argument"\,\cite{bramwell}) make the study of the $2D$ XY universality notoriously challenging. \\
\indent  
{\it iii)} In the nearest-neighbours $2D$ XY model, the classical treatment takes advantage of the duality construction\,\cite{savit}, through which one can famously relate the model to the Coulomb gas\,\cite{Minnhagen, Gulacsi1998} or the sine-Gordon model\,\cite{Amit, Gulacsi1998}. However, this is no longer the case already for next-to-nearest-neighbors couplings. \\
\indent
{\it iv)} It is known that in the SR limit, the physics of the $2D$ {\it classical} 
  XY model can be related to the one of the $1D$ {\it quantum}
  XXZ model via its transfer matrix\,\cite{Mattis1984}.\\
 % In the original derivation, one has to neglect the $z-z$ interactions terms with a range $\geq 4$ lattice sites in the resulting Hamiltonian and introduce a suitable, averaged interaction term for distanceslower than three lattice sites\,\cite{Mattis1984}. This allows to introduce a bosonic  hard-core condition and the mapping onto the $1D$ quantum XXZ.
  This approach is based on the mapping to hard-core bosons, and therefore to the XXZ model, and cannot be straightforwardly applied to the the case of $XY$
  LR interactions, as one should show the RG
  irrelevance of terms violating the hard-core condition. Moreover, let us remark that $2D$ boson gas at finite temperature with (isotropic) $1/r^3$ \textit{density-density} interaction do exhibit a BKT transition \cite{Filinov2010}; but this interaction corresponds to a quantum 1D XXZ in which only in the $z-z$ interaction is long range.
  \\
    \indent
{\it v)} Finally, we observe that the treatment of the SR XY model in $2D$ is very much simplified by the introduction of the Villain model, \cite{Villain1975,Jose1977}, %The $2D$ SR Villain model is both an  approximation of the $2D$ SR XY model and a model interesting  {\it per se}. 
which can be mapped exactly onto the Coulomb gas, and shares the same universality class of the SR $XY$ model. The physical reason of their connection in the SR regime, is that the (gapped) amplitude fluctuations of the corresponding $O(2)$ action are irrelevant\,\cite{popov1983functional}. Thus, once the periodic nature of the phase is taken care of, all the relevant information is present in the theory. However, in the LR regime the interplay between amplitude and phase fluctuations cannot be neglected and it is not known whether they still share the same universality class. \\
  %Given the fact that the Villain model is always, at least in principle, solvable, this lack of a mapping between the $XY$ and Villain models with LR interactions  certainly make more difficult the study of the $2D$ LR XY model. 
Despite these difficulties, the study of LR XY model is of great interest: first, since its introduction, the BKT mechanism \cite{berezinsky,KT,K,review} has been found to quantitatively describe the universal scaling appearing in several $2D$ systems with $U(1)$ symmetry, ranging from thin $^{4}$He films\,\cite{Bishop1978} to quasi-2D layered superconductors\,\cite{corson_vanishing_1999, Rendeira_NatPhys07, bilbro_temporal_2011, Lemberger_PRB85_2012, Baity_LB_PRB93_2016},
exciton-polariton systems\,\cite{Nitsche2014}, cold atoms in $2D$ traps\,\cite{dalibard_2006,Murthy_PRL115_2015} and $2D$ electron gases at the interface between insulating oxides in artificial heterostructures\,\cite{reyren_superconducting_2007, Daptary_PhysRevB.94.085104, Monteiro_PhysRevB.96.020504}.
Apart from these experimental realizations, topological defects are expected to be relevant in several natural
phenomena outside the condensed matter realm, such as DNA tangling or
stripe formation\,\cite{Nisoli2014,Seul1991,Stariolo2012}.
To understand how $\sigma_*$ is modified, is then a crucial question in all the cases in which
a LR tail of the interaction can be added or tuned, especially because
the spin-wave interaction term, already present in the SR case, may destroy,
partially or totally, the topological nature of the phase transition. Moreover, the physics of LR interacting systems has recently experienced
a new wave of interest, due to the current experimental realizations on atomic, molecular
and optical (AMO) systems.
In particular, trapped ions\,\cite{monroe2019programmable, britton2012engineered},
Rydberg gases\,\cite{baranov2012condensed} and optical cavities\,\cite{landig2016quantum, Botzung2021}
allowed the observation of plenty of exotic equilibrium and dynamical phenomena induced by
LR interactions, including entanglement and correlations
propagation\,\cite{jurevic2014quasiparticle, richerme2014non},
dynamical phase transitions\,\cite{zhang2017observation, baumann2010dicke},
time crystals\,\cite{rovny2018observation, choi2017observation, zhang2017observation}
and defect scaling\,\cite{safavi2018verification,keesling2019quantum}. These experimental
results stimulated an impressive theoretical activity to characterize the equilibrium and dynamical critical scaling induced by LR interactions in a wide variety of different systems\,\cite{hauke2013spread,  vodola2014kitaev,maghrebi2016causality, gong2016kaleidoscope,defenu2017criticality,defenu2019dynamical,uhrich2020out}.Despite this outpouring theoretical activity and the long-standing relation between topological scaling and LR interactions, the possible corrections induced by power-law decaying couplings to the topological BKT scaling remain an open question, testable in experiments.

\section{Model \& Preliminaries}
\noindent
We consider a system of planar
rotators on a $2D$ lattice of spacing $a$, described by the
Hamiltonian:
\begin{equation} \label{model}
\beta H =  \frac{1}{2} \sum_{\mathbf{i},\mathbf{j}} J_{|\mathbf{i}-\mathbf{j}|} \left[ 1 -  \cos(\theta_{\mathbf{j}} - \theta_\mathbf{i}) \right]
\end{equation}
where $\mathbf{i},\mathbf{j} \in \mathbb{Z}^2$ and
$J_{|\mathbf{i}-\mathbf{j}|}$ has a power-law tail:
$J_{|\mathbf{i}-\mathbf{j}|} \sim \frac{g}{|\mathbf{i}-\mathbf{j}|^{2+ \sigma}}$
for $|\mathbf{i}-\mathbf{j}|\gg1$. The exponent $\sigma$ is assumed
positive in order to ensure additivity of the thermodynamic quantities\,\cite{campa2014physics}. For the following arguments the specific
form of the couplings is not important, as long as that there are no frustration
effects nor competing interactions.

Let us now summarize what we do know for sure about the LR XY model \eqref{model}:\\
\indent
{\it a)} For $\sigma < 2$, at low enough temperatures, the system magnetizes, as rigorously proven in \cite{kunz1976first}. MC simulations at $\sigma=1$ indicate an order-disorder transition and no BKT phase at finite temperature \cite{Romano1988}. Moreover,one could expect that For $\sigma \le 1$ the critical exponents of the ferro-paramagnetic transition are expected to be mean-field \cite{defenu2015fixed}. \\
  \indent
{\it b)} In agreement with {\it a)}, the spin-wave theory in which the cosine is expanded to the quadratic order, {\it without} imposing the periodicity, as in the original Berezinskii calculation\,\cite{berezinsky}, does also magnetize for $\sigma < 2$, since the contribution of the spin fluctuations is of the form $\int d^2q/q^\sigma$ and thus infrared finite.\\
  \indent
  {\it c)} An upper bound for $\sigma_*$ has to be $\sigma_*=2$, i.e.
  for sure there is BKT for $\sigma>2$, as one can deduce even from the Sak's
  argument since $\eta$ is positive. This result is supported by the self-consistent harmonic calculation recently presented in
  \cite{Giachetti21}, which anyway is unable to make even qualitative predictions for $\sigma<2$.

\section{Effective model}
\noindent
We decompose the coupling as $J_{|\mathbf{i}-\mathbf{j}|}=
J^{S}_{|\mathbf{i}-\mathbf{j}|} + g |\mathbf{i}-\mathbf{j}|^{- (2+ \sigma)}$ where
$J^S$ is a SR term taking into account the small-distances behavior of the coupling. At low temperatures, the spin direction varies
smoothly from site to site and, as a consequence, we can expand the SR term
for small phase differences as
$ \cos ( \theta(\mathbf{x}+ \mathbf{r}) -  \theta(\mathbf{x}) ) \sim 1
-  |\nabla \theta|^2/2$. The same, however, it is not automatically
true for the LR term, since far-away pairs, whose phase difference
is not necessarily small, give a significant contribution to the Hamiltonian.

These considerations allow us to write a continuous version of the Hamiltonian in Eq.\,\eqref{model} in terms of the field $\theta(\mathbf{x})$,
namely the Euclidean action
\begin{equation} \label{action}
  S[\theta] =  \frac{J}{2} \int d^{2}x |\nabla \theta|^2 + S_{\mathrm LR},
\end{equation}
where the LR part can be written as 
\begin{equation}
S_{\mathrm LR} = - \frac{g}{2 \gamma_{2,\sigma}} \int d^2 x (\cos \theta \, \nabla^{\sigma} \cos \theta + \sin \theta \,
  \nabla^{\sigma} \sin \theta ),
  \label{Vill}
\end{equation}
with $\gamma_{2, \sigma} = 2^{\sigma} \Gamma(\scriptstyle \frac{1 + \sigma}{2} \displaystyle) \pi^{-1} |\Gamma(\scriptstyle -\frac{\sigma}{2} \displaystyle)|^{-1}$, by using the definition of (bulk) fractional derivative given in Appendix A. The first and the second term in Eq.\,\eqref{action}
account for the short- and long-range
contributions respectively, with $J \sim 1/T$ and $g \sim 1/T$
being the temperature dependent couplings. 
Notice that the result
would be different for a quantum $1D$ chain with LR interactions,
where interactions are still SR along the imaginary time axis \cite{maghrebi2017continuous}.

If $g=0$, by following the usual duality procedure\,\cite{Jose1977},
one can take into account the periodic nature of the field $\theta$ in Eq.\,\eqref{action} by isolating the topological configurations and introducing the vortex fugacity $y = \exp(- \varepsilon_c)$, being $\varepsilon_c$
the corresponding core energy. This, in turn, leads to the Kosteritz-Thouless
RG equations\,\cite{KT,K,Jose1977,LBGiamarchiCastellani} (see \cite{Drouffe,LeBellac} for textbook presentations) which
%\begin{equation} \label{KTRG}
%\begin{split}
%\frac{dJ_{\ell}}{d\ell} &= - 4 \pi^3 J_{\ell}^2 y_{\ell}^2 \\
%\frac{dy_{\ell}}{d\ell} &= (2 - \pi J_{\ell}) y_{\ell}
%\end{split}
%\end{equation}
%where $\ell=\ln(r/a)$ is the logarithmic scale of the problem and
%$r$ represents the RG cutoff-distance. Given the bare action in
%Eq.\,\eqref{action}, the initial conditions of the RG flow read
%$J_{0}=J$ and $y_{0}=y$. Then, the flow equations in Eqs.\,\eqref{KTRG}
feature a line of stable Gaussian fixed points for $y=0$ and
$J > \frac{2}{\pi}$, describing the power-law scaling observed
in the low-temperature BKT phase. For $g$ small enough, we expect to have then a continuum
theory described by the three parameters $J$, $g$ and $y$.

In order to explore the effects of LR interactions, we deform the traditional BKT fixed-points theory with the non-local operator in the second term of Eq.\,\eqref{action}. Since only those fixed-points which are stable under topological perturbation correspond to an infra-red (IR) limit of the SR BKT theory, we can restrict ourselves to the region in which the topological excitations are
irrelevant ($J>\frac{2}{\pi}$). The relevance of the LR perturbation depends on the scaling dimension
$\Delta_{g}$ of the coupling $g$, which is defined according
to the asymptotic behavior $g_{\ell}\approx\exp(\Delta_{g}\ell)$ for $\ell\gg1$,
where as usual in the BKT literature, we put 
$\ell=\ln(r/a)$. On the other hand, due to the Gaussian nature of the measure,
\begin{equation}
\label{lr_scaling} 
\me{\cos \left( \theta(\mathbf{x}) - \theta(\mathbf{x'}) \right)} = e^{- \frac{1}{2} \me{\left( \theta(\mathbf{x}) - \theta(\mathbf{x'}) \right)^2}} = |\mathbf{x} - \mathbf{x'}|^{-\eta_{\rm sr}(J)},  
\end{equation}
where $\eta_{\rm sr}(J) = \frac{1}{2 \pi J}$ is the exponent of the SR two-point function, \cite{KT,K,Jose1977}.
Following Eq.\,\eqref{lr_scaling}, the scaling dimension of the LR term reads
\begin{align} \label{ScalingDimension}
\Delta_{g}=2-\sigma-\eta_{\mathrm{sr}}(J)
\end{align}
so that the LR perturbation becomes relevant only if
$\sigma < 2-\eta_{\rm sr}(J)$, similarly to the
traditional spontaneous symmetry breaking (SSB) case\,\cite{defenu2015fixed}, but with a
temperature-dependent anomalous dimension. This confirms that
for $\sigma > 2$ the LR perturbation is always irrelevant, as expected.

Let us now consider the case $\sigma < 2$. There, the LR perturbation
becomes relevant at small temperatures, since $\eta_{\rm sr}\simeq 0$ for
$T\simeq 0$. Since $\eta_{sr}$ in Eq.\,\eqref{ScalingDimension} is the one of the SR unperturbed theory, we can apply the results of the traditional BKT theory\,\cite{jose1977renormalization} as long as the LR perturbation is not relevant. In particular we know that topologicaly excitations are irrelevant for $\eta_{sr} < 1/4$, so that in the range $7/4 <\sigma < 2$, a subset of the BKT fixed points remains stable and we have quasi-long-range order (qlro) for a certain temperature window.
This result is rather %extraordinary,
non trivial, since in SSB transitions the traditional Sak's result\,\cite{sak1973recursion} predicts the irrelevance of LR couplings at
all temperatures for $\sigma>2-\eta_{\rm sr}$.

\section{RG Flow} 
\noindent
These results may be %made rigorous
confirmed by deriving the flow equations for the LR
term at the leading order in $g$ for $y=0$, obtaining (see Appendix B): 
\begin{equation} \label{RG LR}
\begin{split}
\frac{d g_{\ell}}{d\ell} &=\Big( 2 - \sigma - \eta_{\rm sr}(J_{\ell}) \Big) g_{\ell}  \\
\frac{dJ_{\ell}}{d\ell} &= c_{\sigma} \eta_{sr}(J_{\ell}) g_{\ell} 
\end{split}
\end{equation}
where $c_{\sigma} = \frac{\pi}{2} a^{2-\sigma} \int^{\infty}_1 du \ u^{1 - \sigma} \mathcal{J}_{0}
(2 \pi u)$ and $\mathcal{J}_{0} (x)$ is the zeroth order Bessel function
of the first kind. As shown in Appendix B the above result is reliable as long as $a^{2-\sigma} g_{\ell} \ll J_{\ell}$ or, equivalently, as long as $\frac{dJ}{d\ell} \ll J_{\ell}$. As expected, we see that the flow equations\,\eqref{RG LR} support a
line of SR fixed points $g=0$ which becomes unstable for
$\eta_{\rm sr}(J)<2-\sigma$. As long as our hypothesis of small $g$ holds, we can  explicitly
identify the form of the flow trajectories of Eqs.\,\eqref{RG LR}: 
\begin{equation} \label{flux}
g_{\ell} (J) =  \frac{\pi (2 - \sigma)}{ c_{\sigma}} \left[ \left( J_{\ell} - J_{\sigma} \right)^2 + k \right],
\end{equation}
where $k$ is a real number and $J_{\sigma} = \frac{1}{2 \pi (2-\sigma)}$.
The sign of $k$ divides the trajectories which met the fixed point $g=0$ and those which do not, the first ones ending at (starting from) the fixed point line for $J\leq J_{\sigma}$\,($J>J_{\sigma}$).
The separatrix corresponds to the semi-parabola with $k=0$, $ J \leq J_{\sigma}$. For $k>0$ $g \to\infty$, showing the existence of a new low-temperature phase, where LR interactions are relevant.
The critical temperature $T_c$, below which this new phase appears, is such that $\eta_{\rm sr}(J_c) > 2-\sigma$ .

Since, as in the traditional BKT calculation\,\cite{KT}, Eqs.\,\eqref{RG LR} were derived for small $g$ and $y$, its use for $T<T_c$ is in principle not justified,
since LR interactions are relevant and $g_{\ell}$ grows indefinitely. However, let us notice that the scaling of $g_{\ell}$ with $T$ for $T \rightarrow T_c^{-}$ can be reliably predicted from Eqs.\,\eqref{RG LR}, since in this limit the flow spends a divergent amount of RG time $\ell$ in the vicinity of the line of fixed points $g=0$. This scaling is derived in Appendix C. Moreover, we can guess the infrared form of the action in the low temperature phase by observing that the rigorous result of Ref.\,\cite{kunz1976first}
implies that for $T<T_c$ the system displays finite magnetization and, then, phase fluctuations are limited even at large
distances. Therefore, the expansion of the
trigonometric function in Eq.\,\eqref{action} is justified leading to an action of the form
\begin{equation} \label{Gaussian}
S_g = - \frac{\bar{g}}{2} \int d^2 x \ \theta \nabla^{\sigma} \theta,
\end{equation}
%where $\bar{g}\sim g_{\ell}$ at large $\ell$.
%for some $\bar{g}$.
where $\bar{g} = g \gamma_{2,\sigma}^{-1}$. Being the above action quadratic, the properties of this exotic low temperature phase can be worked out: in particular the scaling of the magnetization for $T \rightarrow T_c^{-}$ is found to be (see Appendix C for details) 
\begin{equation} \label{scaling}
\ln m \sim -e^{B(T_c - T)^{-1/2}}
\end{equation}
where $B$ is a non universal constant. %It is interesting to see that
Since all the derivatives of $m$ with respect to $T$ vanish at $T=T_c$ (essential singularity), and since $m$ is linked to the derivative of the free energy with respect to the external field, we have that the phase transition between the ordered and disordered
phase is actually of infinite order. Moreover,
the connected correlation functions have a power-law decay for
$T < T_c$ given by $\me{\mathbf{S}(\mathbf{r}) \cdot \mathbf{S}(\mathbf{0})}_{c} \sim \frac{1}{r^{2 - \sigma}}$, where $\mathbf{S}(\mathbf{r})=
  (\cos{\theta}_\mathbf{r},\sin{\theta}_\mathbf{r})$.
\begin{figure}
    \centering
    \includegraphics[scale=0.44]{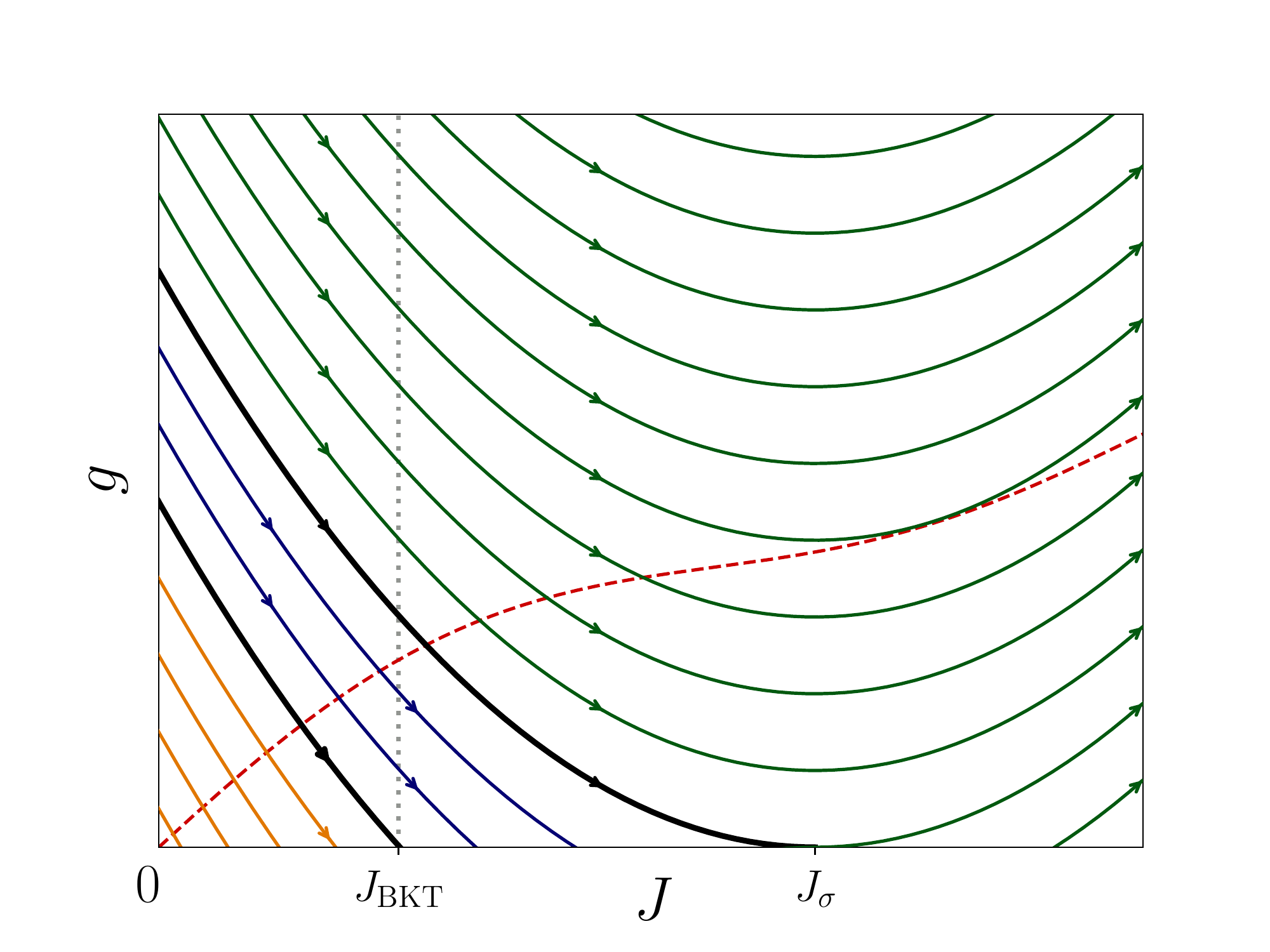}
    \caption{Sketch of the RG flow lines for $\frac{7}{4} < \sigma < 2$
      in the $y=0$ plane. The dashed red line is a possible realization
      of the physical parameters line, from which the flow starts,
      as the temperature is varied. On the right/left of the gray
      dotted line the vortex fugacity $y$ is irrelevant/relevant
      ($\dot{y}_{\ell}/y_{\ell} \gtrless 0 $). The two separatrices
      (bold black lines) divide the flow in three regions: a high-temperature region (orange, the flow ends up in the disordered phase), an intermediate one (blue, the flow reaches a $g=0$ fixed point) and the low-temperature region (green, the LR perturbation brings the system away from the critical line).}
    \label{Fig1}
\end{figure}

We have so far assumed $y=0$; let us now consider the effect of topological excitations. At leading order in both $g$ and $y$ the two perturbations remain independent and, since the vortices contribute to the $J_{\ell}$ flow only beyond leading order in $y$, Eqs.\eqref{RG LR} are unchanged. Moreover, one has $\frac{dy_{\ell}}{d\ell} = (2 - \pi J_{\ell}) y_{\ell}$ valid up to second order terms in $y_{\ell}$ and $g_{\ell}$. Then, in agreement with what we stated above, as long as $\frac{7}{4} < \sigma < 2$, the temperature
range $T$ between $T_c$ and $T_{\rm BKT}$ of
the line of fixed points $g=y=0$ remains stable under
both topological and LR perturbations. In the low-temperature phase instead, it is natural to suppose $y$ to be irrelevant, due to the fact that a non-negligible LR coupling increases the cost of, highly non-local topological excitations. This idea is made rigorous in Appendix D where the interaction energy between vortices in the low temperature phase is computed, and it is shown that they cannot proliferate.

\begin{figure}
    \centering
    \includegraphics[scale=0.34]{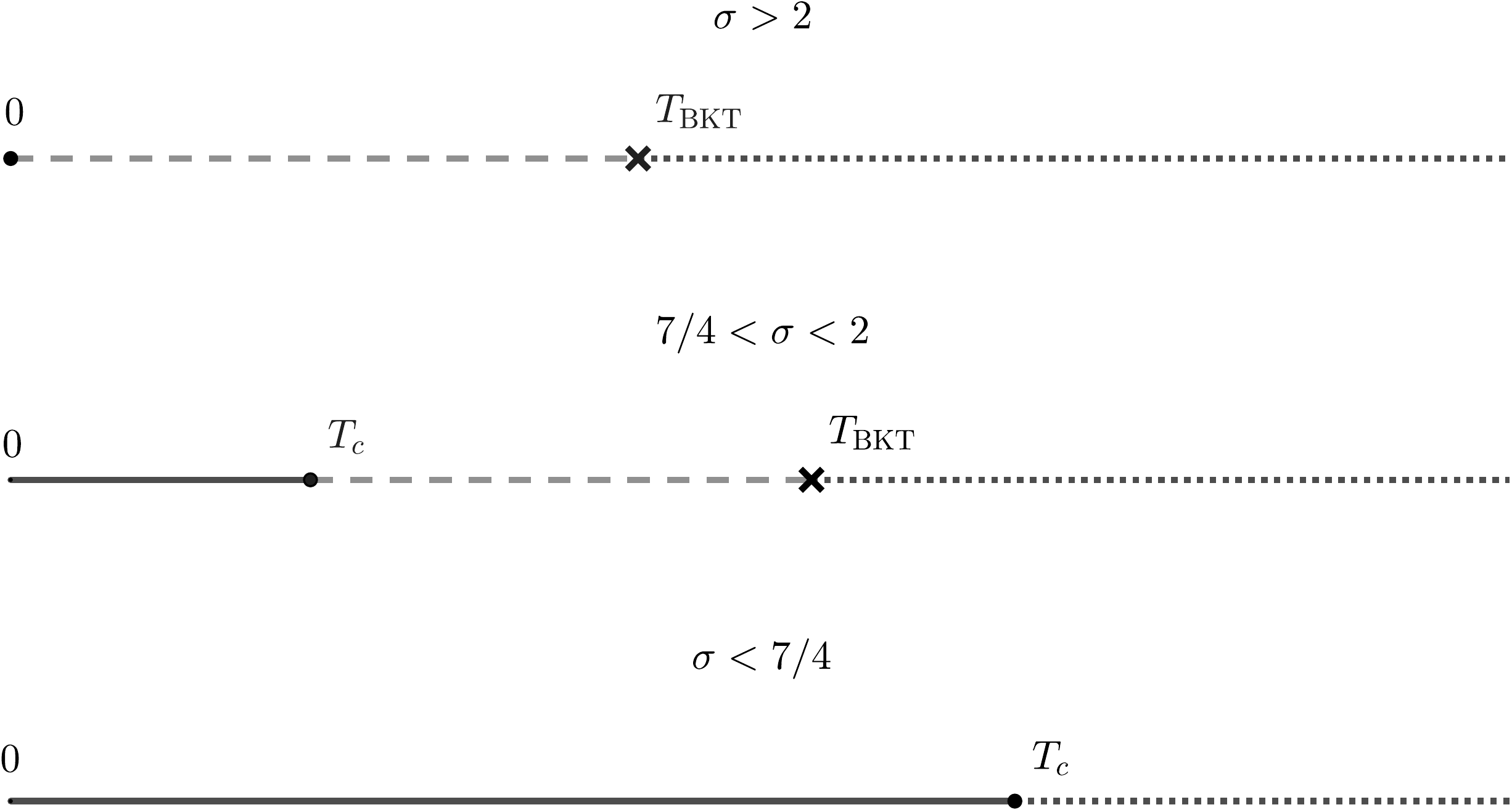}
    \caption{Sketch of the possible phases of the model: ordered with
      magnetization (solid black), BKT qlro
      (dashed light gray), disordered (dashed dark gray).
      If $\sigma>2$ we find the usual SR phenomenology with a
      BKT phase transition. For $\sigma < 2$ an ordered phase
      appears at low-temperatures, the BKT qlro phase
      disappearing for $\sigma < \frac{7}{4}$.}
    \label{Fig2}
\end{figure}

Summarizing, for $\sigma \in (7/4,2)$ we find three phases: \textit{i)} an ordered phase for $T<T_c$ with finite magnetization and temperature independent
power-law correlation functions \textit{ii)} an intermediate BKT phase for $T_c<T< T_{\rm BKT}$, where the magnetization vanishes
and the exponent of the two-point correlation function
depends on $T$ \textit{iii)} a disordered phase for $T>T_{\rm BKT}$. Due to the LR character
of the interactions, also the high-temperature phase displays
power-law decaying two-point functions $\me{\mathbf{S}(\mathbf{r}) \cdot \mathbf{S} (0) } \sim r^{-2-\sigma}$,\cite{spohn1999decay,kargol2005decay,kargol2014decay}.
As $\sigma \rightarrow 7/4^{+}$ the critical temperature $T_c$ reaches $T_{BKT}$ from below. Therefore, for $\sigma < 7/4$, the whole BKT line fixed points becomes unstable either with respect to topological or LR perturbations and the intermediate phase vanishes, leaving only a SSB phase transition. However, our approach cannot reliably be used to fully characterize this transition: as $T$ approaches $T_c$ from below, the RG flow slows down close to the $g=0$, $J = J_{\sigma}$ fixed point. Since for $\sigma < \frac{7}{4}$ $J_{\sigma} < J_{\rm BKT}$, $y$ grows indefinitely, away from the $y\ll1$ regime. Our results are summarized in Fig.\,\ref{Fig2}.

\section*{Conclusions}
\noindent
We have shown that the introduction of long-range (LR)
power decaying couplings in the $2D$ XY model Hamiltonian
produces a 
rich phase diagram, different from the short-range
(SR) case\,\cite{KT} and from the one of $O(N)$ LR systems\,\cite{sak1973recursion}. Remarkably, for $7/4 < \sigma < 2$, the system displays both BKT qlro in the temperature interval $T_c < T < T_{\rm BKT}$ and actual long-range order
for $T<T_c$.

The introduction of complex interaction patterns in systems with $U(1)$
symmetry is known to generate exotic critical features, as in the anisotropic
$3D$ XY model\,\cite{shenoy1995anisotropic},
coupled XY planes\,\cite{bighin2019berezinskii}, 2D systems with anisotropic
  dipolar interactions\,\cite{Maier2004,Fischer2014} or four-body interactions\,\cite{Antenucci2015}, 
and high-dimensional systems with Lifshitz
criticality\,\cite{jacobs1983self,defenu2020topological}
%\textcolor{red}{on the other hand BKT phenomenology is known to survive in presence of dipolar interaction\,\cite{Fischer2014}}.
The present work constitutes a further milestone along this path, as it identifies a peculiar critical behavior, namely a non-analytic exponential vanishing of the order parameter, that eludes the current classification of universal scaling behaviors\,\cite{raju2019normal}.

Our predictions may be tested in several low dimensional AMO systems. It would be interesting to perform extensive numerical simulations in order to observe the scaling of the critical quantities, and especially the magnetization, close to the low-temperature endpoint of the BKT line in the regime $7/4<\sigma<2$. These simulations will be useful to 
classify this unprecedented transition and to investigate possible corrections
near the $\sigma=7/4$ endpoint due to higher-order
effects caused by spin-wave excitations\,\cite{Maccari2020}.
Further investigation is also needed to compare our
results with the LR diluted model studied in\,\cite{leuzzi2013,cescatti2019analysis}. In this model, at $\sigma=1.875$, the
numerical simulations presented in \,\cite{cescatti2019analysis} 
do not
find any intermediate BKT region, but the general question 
whether the $2D$ LR diluted XY model and the $2D$ LR non-diluted one have the
same phase diagram
remains open. 

Our results have also implications for LR quantum
XXZ chains \cite{maghrebi2017continuous,bermudez,botzung2019effects}.
One would need to perform the exact mapping of the classical
$2D$ LR XY model to an effective $1D$ quantum model, following the calculation
presented in \cite{Mattis1984} and valid for the classical
$2D$ SR XY model. If the non-local/LR terms violating the hard-core boson
condition can be shown to be irrelevant,
then one could put in correspondence
our phase diagram with that of the LR quantum
XXZ chains having LR couplings both for $x-y$ and $z-z$ terms
\cite{maghrebi2017continuous}. This seems to be confirmed
by the similar structure of the RG flow equations
of \cite{maghrebi2017continuous} with our Eqs.\eqref{RG LR}
taken at low temperatures.
If this is the case, then the two lines,
black and white, of Fig.1 of \cite{maghrebi2017continuous}
would merge in a point,
with the XY phase disappearing, corresponding to our $\sigma=7/4$ point. Finally, we mention that it would be interesting to study in detail the phase diagram of the $2D$ LR Villain model for $\sigma<2$.

\section{Acknowledgment}
\noindent
Valuable discussions with G. Parisi, F. Ricci-Tersenghi and A. Scardicchio are gratefully acknowledged. N.D. and A.T. also acknowledges useful discussion with M. Ib\'a\~nez Berganza. This work is supported by the Deutsche Forschungsgemeinschaft (DFG, German Research Foundation) under Germany’s Excellence Strategy EXC2181/1-390900948 (the Heidelberg STRUCTURES Excellence Cluster). This work is supported by the CNR/HAS (Italy-Hungary) project
“Strongly interacting systems in confined geometries". This work is part of MUR-PRIN2017 project “Coarse-grained description for non-equilibrium systems and transport phenomena (CONEST)" No. 201798CZL whose partial financial support
is acknowledged.

\section{Appendix}
\noindent
\subsection{A. Defintion of the fractional Laplacian}
\noindent
Given a real parameter $\sigma \in (0,2)$,
one can define the fractional Laplacian of order $\sigma$ of a function
$f(\mathbf{x}): \, \mathbb{R}^d \rightarrow \mathbb{R}$ as:
\begin{equation} \label{frac}
\nabla^{\sigma} f(\mathbf{x}) \equiv \gamma_{d,\sigma} \int d^d r \frac{f(\mathbf{x}+\mathbf{r}) - f(\mathbf{x}) }{r^{d + \sigma}},    
\end{equation}
where $\gamma_{d,\sigma} =  \frac{2^{\sigma} \Gamma(\frac{d+ \sigma}{2})}{\pi^{d/2} | \Gamma(- \frac{\sigma}{2})|}$ and $r=\mid \mathbf{r} \mid$.
Another expression for this quantity can be derived in terms of the Fourier transform, $f(\mathbf{q})$, of $f(\mathbf{x})$:
\begin{equation}
\nabla^{\sigma} f (\mathbf{x}) = - \gamma_{d, \sigma} \int d^d q \ f(\mathbf{q}) \ e^{i \mathbf{q} \cdot \mathbf{x}} \int d^d r \frac{1 - e^{i \mathbf{q} \cdot \mathbf{r}}}{r^{d+ \sigma}}.
\end{equation}
Since 
\begin{equation}
\int d^d r \frac{1 - e^{i \mathbf{q} \cdot \mathbf{r}}}{r^{d+ \sigma}} = \gamma_{d,\sigma}^{-1} \ q^{\sigma},
\end{equation}
we find the alternative definition:
\begin{equation}
\nabla^{\sigma} f (\mathbf{x}) = - \int d^d q \ q^{\sigma} f(\mathbf{q}) e^{i \mathbf{q} \cdot \mathbf{x}}. 
\end{equation}
%(which actually justifies the name of the operator). \\

In our case $d=2$ and we have to evaluate the quantity: 
\begin{equation}
\int d^2 x \int_{r>a} \frac{d^2 r}{r^{2+ \sigma}} [1 - \cos \left(\theta (\mathbf{x}) - \theta (\mathbf{x} + \mathbf{r} ) \right)]
\end{equation}
For $\sigma < 2$, one can actually disregard the contribution
coming from the lattice spacing $a$, since it would just result in a
correction of the short-range term. Then,
through trivial trigonometric manipulations we can write the above expression
as: 
\begin{equation}
\begin{split}
& \int d^2 x \cos \theta (\mathbf{x}) \int \frac{d^2 r}{r^{2+ \sigma}} [ \cos \theta (\mathbf{x}) - \cos \theta (\mathbf{x} + \mathbf{r} ) ]  \\ +
& \int d^2 x \sin \theta (\mathbf{x}) \int \frac{d^2 r}{r^{2+ \sigma}} [ \sin \theta (\mathbf{x}) - \sin \theta (\mathbf{x} + \mathbf{r} ) ].
\end{split}
\end{equation}
Finally, using the definition \eqref{frac} of the fractional derivative,
we get
\begin{equation}
- \gamma_{2,\sigma}^{-1} \int d^2 x \left( \cos \theta \nabla^{\sigma} \cos \theta + \sin \theta \nabla^{\sigma} \sin \theta \right)
\end{equation}
which justifies the alternative form of the long-range term
given in the main text as Eq.(4).%\eqref{Vill}

\subsection{B. Renormalization group for $y=0$}
\noindent
We will now derive the set of RG equations (7) %\eqref{RG LR}
given in the main text,
valid for $y=0$. %Since in this case no topological effect is presentù and we can then ignore the constraint in equation \eqref{Z} of the main text.
We then start form the action written in the form 
\begin{equation} \label{StartingAction}
S[\theta] = \int d^2 x \left( \frac{J_{\ell}}{2} | \nabla \theta|^2 + \frac{g_{\ell}}{2} \int_{r>a} \frac{d^2 r}{r^{2+ \sigma}} \left[ 1 - \cos \left( \Delta_{\mathbf{r}} \theta (\mathbf{x}) \right) \right] \right)
\end{equation}
where, as in the main text, $\Delta_{\mathbf{r}} \theta(\mathbf{x}) = \theta(\mathbf{x}+\mathbf{r})-\theta(\mathbf{x})$, and compute the flux perturbatively around $g=0$.
%We use a wilsonian renormalization scheme:
The field is split into fast and slow modes with respect to the
momentum cutoff $\Lambda = \frac{2 \pi}{a}$, namely
$\theta = \theta^{>} + \theta^{<}$ with
\begin{equation}
\begin{split}
\theta^{<} (\mathbf{x}) &= \int_{q < \Lambda e^{- d \ell}} \frac{d^2 q}{(2 \pi)^2} \theta(\mathbf{q}) e^{i \mathbf{q} \cdot \mathbf{x}} \\
\theta^{>} (\mathbf{x}) &= \int_{\Lambda > q > \Lambda e^{- d \ell}} \frac{d^2 q}{(2 \pi)^2} \theta(\mathbf{q}) e^{i \mathbf{q} \cdot \mathbf{x}},
\end{split}
\end{equation}
where $\ell=\ln(r/a)$. If we assume the interacting long-range term in Eq.\,\eqref{StartingAction} to be small with respect to the quadratic one, we can perform the integration perturbatively. It is easy to see that this is possible if $g a^{2-\sigma} << J$. Under this assumption then we integrate out the fast modes, expanding the partition function in cumulants of the non-Gaussian part $S_g$: 
\begin{equation}
S_{\rm eff} [\theta^{<}] = S_0 [\theta^{<}] +  \me{S_g}_{>}  + O(g^2). 
\end{equation} 
Writing $\cos(\Delta_{\mathbf{r}} \theta) =
\cos(\Delta_{\mathbf{r}} \theta^{>}) \cos(\Delta_{\mathbf{r}} \theta^{<}) +
\sin(\Delta_{\mathbf{r}} \theta^{>}) \sin(\Delta_{\mathbf{r}} \theta^{<})$,
one sees that only the first term will give a contribution.
Then, up to additive constants we have:
\begin{equation}
\me{S_g}_{>} =  \frac{g_{\ell}}{2} \int d^2 x  \int \frac{d^2 r}{r^{2+ \sigma}} \me{\cos(\Delta_{\mathbf{r}} \theta^{>})}_{>} \left[ 1 - \cos \left( \Delta_{\mathbf{r}} \theta^{<}  \right) \right]
\end{equation}
(from now on we omit the $r>a$ condition in the integral over $r$).
On the other hand, $\me{\cos(\Delta_{\mathbf{r}} \theta^{>})}_{>} = e^{- \frac{1}{2} \me{\left( \theta(\mathbf{r})- \theta(0) \right)^2}_{>}}$ and
\begin{equation}
%\begin{split}
  \frac{1}{2} %&
  \me{\left( \theta(\mathbf{r})- \theta(0) \right)^2}_{>}  %\\ &
  = \int_{\Lambda > q > \Lambda e^{- d \ell}} \frac{d^2 q}{(2 \pi)^2} \frac{1 - \cos(\mathbf{q} \cdot \mathbf{r})}{J_{\ell} q^2}  %\\ &
  = \frac{d \ell}{2 \pi J_{\ell}} \Big(1 - \mathcal{J}_0(\Lambda r)\Big), 
%\end{split}
\end{equation}
where $\mathcal{J}_0 (x)$ is the zeroth-order Bessel function of the first kind. Then,
introducing $\eta_{\rm sr} (J) = \frac{1}{2 \pi J_\ell}$, the exponent of the correlations at the cutoff scale $\ell$, we have:
\begin{equation}
\begin{split}
\me{ \cos(\Delta_{\mathbf{r}} \theta^{>})}_{>}  &= e^{- \eta_{\rm sr}(J_\ell) d \ell \left( 1 - \mathcal{J}_0 (\Lambda r) \right)}  \\ &= 1 -  \eta_{\rm sr}(J_\ell) d \ell + \eta_{\rm sr}(J_\ell) d \ell \mathcal{J}_0(\Lambda r)
\end{split}
\end{equation}
up to second order corrections. The first two terms provide
an anomalous dimension of the coupling
$g_{\ell + d \ell} = g_{\ell} e^{- \eta_{\rm sr} (J_\ell) d \ell}$, as expected,
while the last one modifies the power-law dependence
on $r$ of the long-range term: 
\begin{equation}
\begin{split}
  \me{S_g}_{>} = & \frac{1}{2} \int d^2 x \Biggl\{ \int \frac{d^2 r}{r^{2+ \sigma}}  g e^{- \eta_{\rm sr}
    (J_\ell) d \ell}  \left[ 1 - \cos \left( \Delta_{\mathbf{r}} \theta^{<} \right) \right] \\
 + &g \eta_{\rm sr}(J_\ell) d \ell   \int \frac{d^2 r}{r^{2+ \sigma}} \mathcal{J}_0(\Lambda r) \left[ 1 - \cos \left( \Delta_{\mathbf{r}} \theta^{<} \right) \right]  \Biggr\}.
\end{split}
\end{equation}

Let us now examine the last term of the above equation.
This can be seen as an interaction term of the original $XY$ form.
Since $\mathcal{J}_0 (x) \sim x^{-1/2} \cos(x - \pi/4)$ for large $x$,
the new coupling decays faster than the original and has
an oscillating behavior, which provides a natural cutoff for $r \sim \Lambda^{-1}$. It is then reasonable
to approximate this with a short-range coupling of the form
$|\nabla \theta|^2$. The simplest way is to replace
$1 - \cos(\Delta_{r} \theta) \approx
\frac{1}{2}(\mathbf{r} \cdot \nabla_{\mathbf{x}} \theta)^2$ and observe that
\begin{equation}
\int \frac{d^2 r}{r^{2+ \sigma}} \mathcal{J}_0(\Lambda r) (\mathbf{r} \cdot \nabla_{\mathbf{x}} \theta^{<})^2 = \pi |\nabla_{\mathbf{x}} \theta^{<}|^2 \int^{\Lambda^{-1}}_{a} dr r^{1- \sigma} \mathcal{J}_0(\Lambda r) .
\end{equation}
For $\sigma > \frac{1}{2}$, we can neglect the cutoff and, with the substitution $r = a u $, we can express the correction in the action as 
\begin{equation}
\frac{c_{\sigma}}{2}  (g_{\ell} a^{2- \sigma}) \eta_{\rm sr} (J_\ell) d \ell \int d^2 x |\nabla_{\mathbf{x}} \theta^{<}|^2,
\end{equation}
where
$c_{\sigma} = \frac{\pi}{2} \int_1^{\infty} du
u^{1-\sigma} \mathcal{J}_0(2 \pi u) > 0 $. The integral is actually
ill-defined for $\sigma < \frac{1}{2}$ signaling that our approximation breaks down. %only the first oscillations have to be taken into account.
Let us notice however that the precise value of the coefficient
is not important for our analysis. Moreover, it should be noticed that this entire procedure is only reliable for $\sigma > 7/4$, where part of the BKT fixed points line remains stable and furnishes a viable expansion point, see the discussion in the  main text.
Up to the first order in $g$, then the integration of the fast modes
gives the corrections: 
\begin{equation}
\begin{split}
dg &= - \eta_{\rm sr} (J_{\ell}) g_{\ell} d \ell \\
dJ &= c_{\sigma} \eta_{\rm sr} (J_{\ell}) (g_{\ell} a^{2- \sigma})   d \ell.
\end{split}
\end{equation}

In order to obtain a theory with the same cutoff scale, we have
to do the replacement $\mathbf{x} \rightarrow \mathbf{x} e^{-d \ell}$ in
the action.
%change the cutoff in the action $a \rightarrow a e^{-d \ell}$. 
This modifies the couplings $g$, $J$ by their own bare length dimension,
i.e. $2 - \sigma$ and $0$ respectively:
\begin{equation}
\begin{split}
dg &= (2 - \sigma - \eta_{\rm sr} (J_{\ell}) ) g_{\ell} d \ell \\
dJ &= c_{\sigma} \eta_{\rm sr} (J_{\ell}) (g_{\ell} a_0^{2- \sigma})   d \ell.
\end{split}
\end{equation}
In turn, one finally obtains the RG equations:
\begin{equation} \label{RG LR}
\begin{split}
\frac{dg}{d\ell} &=\left(2 - \sigma - \eta_{\rm sr} (J_{\ell}) \right) g_{\ell}  \\
\frac{dJ}{d\ell} &= c_{\sigma} \eta_{\rm sr} (J_{\ell}) g_{\ell}, 
\end{split}
\end{equation}
i.e. Eqs. (7) %\eqref{RG LRx}
of the main text (we absorbed the constant ultraviolet cutoff $a^{2 - \sigma}$
in the definition of $c_{\sigma}$). 

\subsection{C. Magnetization in the low-temperature phase}
\noindent
We will now derive the scaling behavior (10) %\eqref{scaling}
given in the main text for the magnetization near $T_c$, for $ T \to T_c^{-}$.
We start from the Gaussian theory, %\eqref{Gaussian}
Eq. (9) of the main text, describing
the low temperature phase of the theory in the infrared (IR).
Being the theory Gaussian, it is
$m = \me{\cos \theta (\mathbf{x})} = e^{- \frac{1}{2} \me{\theta^2(\mathbf{x)}}}$.
Being
\begin{equation}
\me{\theta^2(\mathbf{x})} = \int_{q < 2 \pi/a} \frac{d^2 q}{(2 \pi^2)} \frac{1}{ \bar{g} q^{\sigma}} \sim \frac{1}{\bar{g} a^{\sigma - 2}}.
\end{equation}
we find
\begin{equation} \label{mofg}
m = e^{- A/\bar{g}},
\end{equation} 
where $A$ is a non-universal constant. Now, from the flow equations \eqref{RG LR}, we find: 
\begin{equation}
g_{\ell} = g e^{(2-\sigma) \ell} e^{- \int \eta_{\rm sr} (J_{\ell}) d \ell},  
\end{equation}
which is reliable as long as $g_{\ell}$ is small. Let us consider a trajectory which runs very close to the separatrix which, according to Eq. (8) of the main text, is described by the trajectory $g = \frac{\pi (2 - \sigma)}{c_{\sigma}} \left[ (J- J_{\sigma})^2 + k \right]$ with $k \rightarrow 0^{+}$. Let us consider a point in the flow $\ell^{*}$ such that $g(\ell^{*})$ is small and $J(\ell^{*}) > J_{\sigma}$. Then: 
\begin{equation}
\int^{\ell^{*}}_0 \eta_{\rm sr} (J_{\ell}) d \ell = \int^{\ell^{*}}_{J_0} \eta_{\rm sr} (J) \ \frac{dJ}{\dot{J}} = c_{\sigma}^{-1} \int^{\ell^{*}}_{J_0} \frac{dJ}{g(J)} = \pi(2 - \sigma) \int^{J(\ell^{*})}_{J_0} \frac{dJ}{ \left( J - J_{\sigma}	\right)^2 + k }
\end{equation}
By changing the value of the temperature, we have that $J_0$ 
%If we decrease the temperature,
crosses the separatrix ($k \rightarrow 0^{+}$) for some $J_c < J_{\sigma}$ that
corresponds to the critical temperature $T_c$, and consequently $k \sim T_c - T$. Since in this case the
integration interval on $J$ contains the second order singularity $J_{\sigma}$, we have that the integral diverges as $k^{-1/2}$ as $k \rightarrow 0^{+}$. Then we have
\begin{equation}
    g_{\ell^{*}} \sim e^{-B(T-T_c)^{-1/2}}
\end{equation}
where $B$ is a non universal constant. Since, as $k \rightarrow 0^{+}$, the trajectories corresponding to different values of $k$ run close in the parameter space, for large $g$ as well, we do not expect this scaling to be modified in the non-perturbative region. Finally, exploiting Eq.\,\eqref{mofg}, one has the scaling:
\begin{equation}
    \ln m \sim - A e^{B(T-T_c)^{-1/2}}
\end{equation}

\subsection{D. Irrelevance of topological excitations in the low-temperature phase}
\noindent
We start from the quadratic action of Eq. (9) of the main text, which describes the low temperature phase, we express it in terms of the Fourier transform of $\mathbf{v}(\mathbf{x}) = \nabla \theta$ 
\begin{equation} 
S_g = \bar{g} \int \frac{d^2 q}{(2 \pi)^2} \ q^{\sigma} \ |\mathbf{\theta} (\mathbf{q})|^2 = \bar{g} \int \frac{d^2 q}{(2 \pi)^2} \ q^{\sigma - 2} \ |\mathbf{v} (\mathbf{q})|^2
\end{equation}
We notice that circling around a topologically non-trivial region we have 
\begin{equation}
\oint \nabla \theta \cdot d \mathbf{r} = \oint \mathbf{v} \cdot d \mathbf{r} = 2 \pi m_{enc}
\end{equation}
were $m_{enc}$ is the sum of all the topological charges $m_i$ enclosed in the integration contour. This can be rephrased by saying that $ \nabla \times \mathbf{v} (\mathbf{x})= 2 \pi n(\mathbf{x})$, where $n(\mathbf{x}) = \sum_j m_j \delta(x-x_j)$ is the vortex-density and $x_j$ correspond to the positions of the vortices. This can be further simplified if we introduce the dual $\mathbf{u} (\mathbf{x})$ of $\mathbf{v}(\mathbf{x})$, defined as $u_j = \epsilon_{jk} v_k$ where $\epsilon_{jk}$ is the fully antisymmetric tensor of rank $2$. We then find the condition 
\begin{equation}
    \nabla \cdot \mathbf{u} (\mathbf{x}) = 2 \pi n (\mathbf{x})
\end{equation}
In turn, this can be solved in the Fourier space:
\begin{equation}
     \mathbf{u} (\mathbf{q}) = \frac{2 \pi \mathbf{q}}{q^2} n(\mathbf{q)} + \mathbf{u}_{\perp} (\mathbf{q})
\end{equation}
where $\mathbf{u}_{\perp} (\mathbf{q})$ is a generic function such that $\mathbf{q} \cdot \mathbf{u}_{\perp} (\mathbf{q}) = 0$ and which represent the topologically-trivial component of the field $\theta$. Now, since 
\begin{equation}
    |\mathbf{v} (\mathbf{q})|^2 = |\mathbf{u} (\mathbf{q})|^2 = \frac{(2 \pi)^2}{q^2} |n(\mathbf{q})|^2 + |\mathbf{u}_{\perp} (\mathbf{q})|^2
\end{equation}
we have that the action $S_g$ splits into the sum on the non-topological and topological part, the latter being: 
\begin{equation}
    S_{\rm top} = \bar{g} \int d^2 q \ q^{\sigma-4} |n(\mathbf{q})|^2
\end{equation}
Coming back to the real space we have: 
\begin{equation}
    S_{\rm top} = \bar{g}  \sum_{ij} m_i m_j G(\mathbf{r}_i - \mathbf{r}_j)
\end{equation}
with $G(\mathbf{x}) = \int d^2 q \ q^{\sigma-4} e^{i\mathbf{q} \cdot \mathbf{x}} \sim L^{2 - \sigma} - x^{2 - \sigma}$, $L$ being the system size. The first term in $G$ gives raise to a term proportional to $L^{2-\sigma} \sum_{i,j} m_i m_j = L^{2-\sigma} \left( \sum_i m_i \right)^2$ which, in the thermodynamic limit, ensures the neutrality of the gas of charges. We find then
\begin{equation}
    S_{\rm top} \sim - \bar{g}  \sum_{ij} m_i m_j |\mathbf{x}_i - \mathbf{x}_j|^{2-\sigma}
\end{equation}
As expected, this interaction is more binding than the logarithmic one for the short-range case. A simple entropy-energy argument shows that the charges will never unbound at any temperature: indeed the energetic cost of creating two far apart vortices grows like $\bar{g} L^{2-\sigma}$ while the entropy as $\ln L$ so that the free energy
\begin{equation}
F \sim \ln L - \bar{g} T L^{2 - \sigma}
\end{equation}
is always dominated by the interaction term for large enough $L$. 

\bibliography{main}
\end{document}